\documentclass[prb,aps,twocolumn,showpacs,groupedaddress,superscriptaddress]{revtex4}
\usepackage{graphicx,epsf,bm,bbm,color,amsmath,ulem,amssymb}
\newcommand{\beq}{\begin{eqnarray}}
\newcommand{\eeq}{\end{eqnarray}}

\begin{document}

\title{Quantum Hall effect in ac driven graphene: from half-integer to integer case}
\author{Kai-He Ding},
\affiliation{Department of Physics and Electronic Science, Changsha University of Science and Technology, Changsha 410076, P. R. China}
\author{Lih-King Lim}
 \affiliation{Institute for Advanced Study, Tsinghua University, Beijing 100084, P. R. China}
\author{Gang Su}
\affiliation{Theoretical Condensed Matter Physics and Computational Materials Physics Laboratory, School of Physics, University of Chinese Academy of Science, Beijing 100049, P. R. China} 
\author{Zheng-Yu Weng}
\affiliation{Institute for Advanced Study, Tsinghua University, Beijing 100084, P. R. China}
\affiliation{Collaborative Innovation Center of Quantum Matter, Tsinghua University, Beijing 100084, P. R. China}

\begin{abstract}
We theoretically study the quantum Hall effect (QHE) in graphene with an ac electric field. Based on the tight-binding model, the structure of the half-integer Hall plateaus at $\sigma_{xy} = \pm(n + 1/2)4e^2/h$ ($n$ is an integer) gets qualitatively changed with the addition of new integer Hall plateaus at $\sigma_{xy} = \pm n(4e^2/h)$ starting from the edges of the band center regime towards the band center with an increasing ac field. Beyond a critical field strength, a Hall plateau with $\sigma_{xy} = 0$ can be realized at the band center, hence restoring fully a conventional integer QHE with particle-hole symmetry. Within a low-energy Hamiltonian for Dirac cones merging, we show a very good agreement with the tight-binding calculations for the Hall plateau transitions. We also obtain the band structure for driven graphene ribbons to provide a further understanding on the appearance of the new Hall plateaus, showing a trivial insulator behavior for the $\sigma_{xy} = 0$ state. In the presence of disorder, we numerically study the disorder-induced destruction of the quantum Hall states in a finite driven sample and find that qualitative features known in the undriven disordered case are maintained.
\end{abstract}

\pacs{ 73.43.-f, 72.80.Vp, 72.15.Rn, 78.67.Wj}

\maketitle

\section{Introduction}
Graphene exhibits unconventional integer quantum Hall effect (QHE) with the Hall plateaus located at the half-integer positions, i.e., with a Hall conductivity $\sigma_{xy}=\pm (n+1/2)\,(4 e^2/h)$, where $n=0,1,2,\cdots$, usually termed `half-integer' QHE \cite{novoselovnature2005,zhangnature2005,zheng2002,gusynin05,peres06,neto09}. While degeneracy in both the electron spin and valley degree-of-freedom result in a total factor four times the quantum conductance $e^2/h$ across each plateau, the half-integer shift is attributed to the nontrivial Berry phase carried by the quasiparticles. Indeed, the latter is one of the hallmark of Dirac fermions in condensed matter systems. 

For a lattice model of graphene, the system displays a rich interplay between the unconventional and conventional integer QHE for Fermi energy ranging from the band center to the band edge, intercepted by a `crossover' regime with states displaying fluctuating Hall conductivity \cite{shengprb2006,hatsugai06}. In the presence of disorder, it was shown that the unconventional QHE is rather stable against disorder scattering in comparison to the conventional ones, a feature attributed to the Dirac-like quasiparticles \cite{shengprb2006,koshino07,ortmann13,bennaceurprb2015}.  

\begin{figure}[tph]
\centering
\includegraphics[width=0.4\textwidth]{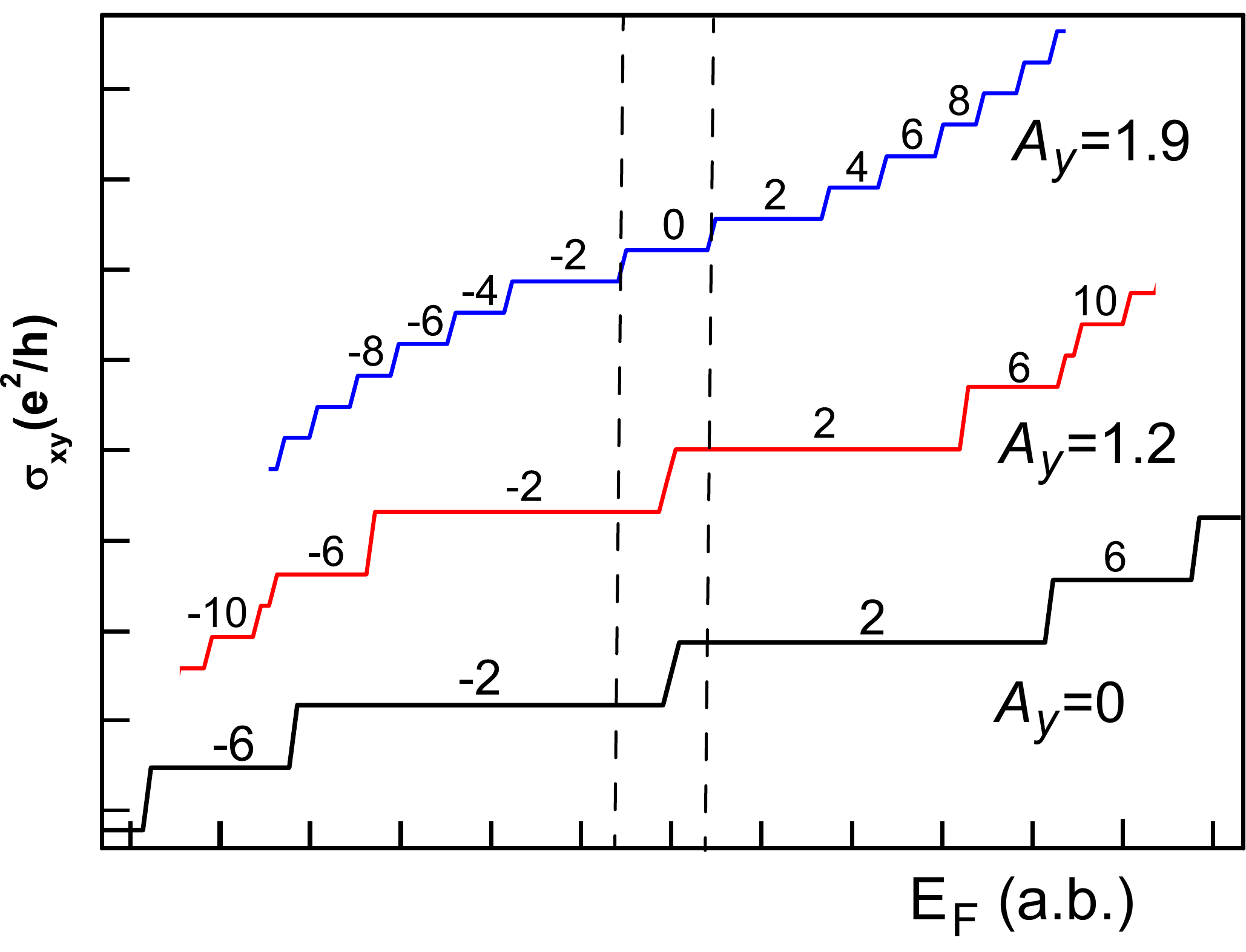}
\caption{Evolution of the Hall conductivity in the band center (region in dash lines) with an increasing ac field strength $A_y$ (at fixed $A_x$) in the high frequency limit. The number on the plateau is the associated Hall conductivity value in unit of $e^2/h$. \label{fig2}}
\end{figure}
It is also known that a time-periodic external field is a powerful tool for the exploration and modification of graphene properties \cite{oka09,kitagawa10,selma12,gomez13,delplace13,agprb2014,kundu14,perez14,wang14,xiong16,plekhanov17}. For example, the external fields can be used as means to control the electronic transmission in a graphene junction \cite{fistulprl2007,dingjpcm2012,gagnon17}, or to drive the system into topological states \cite{oka09}. In this paper, we study the influence of an ac electric field on the quantum Hall effect in graphene lattice model, in clean and disordered scenarios. The main result is summarized in Fig. 1, showing the evolution of the Hall plateaus in a clean driven system. Specifically, the Hall plateau structure evolves in sequence from the original `half-integer' QHE to a complete integer QHE with $\sigma_{xy}= \pm n (2e^2/h)$, as the ac field strength increases. In the fully developed integer QHE regime, a notable feature is the zero Hall plateau at the charge neutrality point. We will see that the application of the ac electric field gradually lifts the valley degeneracy of the Landau levels of graphene \cite{hasegawa06,dietlprl2008,montambauxprb2009}. After the complete lifting of degeneracy, the system at charge neutrality is an insulating state with no edge current, in stark contrast to the undriven graphene where a finite energy density of states persists due to the zero-mode Landau level. In the presence of disorder, the ac field enhances the disorder-induced destruction of the quantum Hall states, further shrinking the quantum Hall parameter regime with respect to the insulating state.

The paper is organized as follows. In Sect. II, we introduce the effective Floquet Hamiltonian for graphene under an ac field. In Sect. III, we study the associated quantum Hall effect. Specifically, we start in Sect. IIIA with the computation of the Hall conductivity. In Sect. IIIB, we utilize a low-energy Hamiltonian with two Dirac cones and compare with the tight-binding Hamiltonian. In Sect. IIIC, we study the Landau level structure in driven graphene ribbons. In Sect. IV, we consider the effect of disorder on the QHE. We end with conclusions in Sect. V.

\section{Graphene under ac field}
Electronic properties of graphene are described by electrons hopping in a honeycomb lattice, which is formed by two interpenetrating triangular A and B sublattices \cite{neto09}. In the presence of an ac electric field, the time-dependent tight-binding Hamiltonian is given by
\begin{equation}
H(\tau)=-\sum\limits_{i\in A}\sum_{m=1}^3 t\, e^{i\phi_{i,\delta_m}(\tau)} a_{i}^\dag \, b_{i+\delta_m} +H.c.,\label{ham}
\end{equation}
where $a_{i}^\dag(a_{i})$ creates (annihilates) an
electron at site $i$ of the sublattice A and $b_{i+\delta_m}^\dag(b_{i+\delta_m})$
creates (annihilates) an electron at site $i+\delta_m$ of the sublattice B, with the connecting vectors
$\delta_1=(a_0/2)(\sqrt{3},1)$, $\delta_2=(a_0/2)(-\sqrt{3},1)$, $\delta_3=a_0(0,-1)$, and $a_0$ is the lattice constant. Here, $\tau$ is the time parameter, $t$ is the nearest-neighbor hopping amplitude, and the ac field is introduced by Peierls substitution via
$\phi_{i,\delta_m}(\tau)=(e/\hbar)\int_i^{i+\delta_m}
\mathbf{A}_{ac}(\tau)\cdot d\mathbf{r}$ with
$\mathbf{A}_{ac}(\tau)=\{\mathcal{A}_x\sin(\omega_0 \tau),
\mathcal{A}_y\sin(\omega_0\tau+\varphi)\}$ representing the ac field with the modulation
frequency $\omega_0$ and $\varphi$ determines the shape of the modulation.
\begin{figure}[tph]
\centering
\includegraphics[width=0.6\columnwidth]{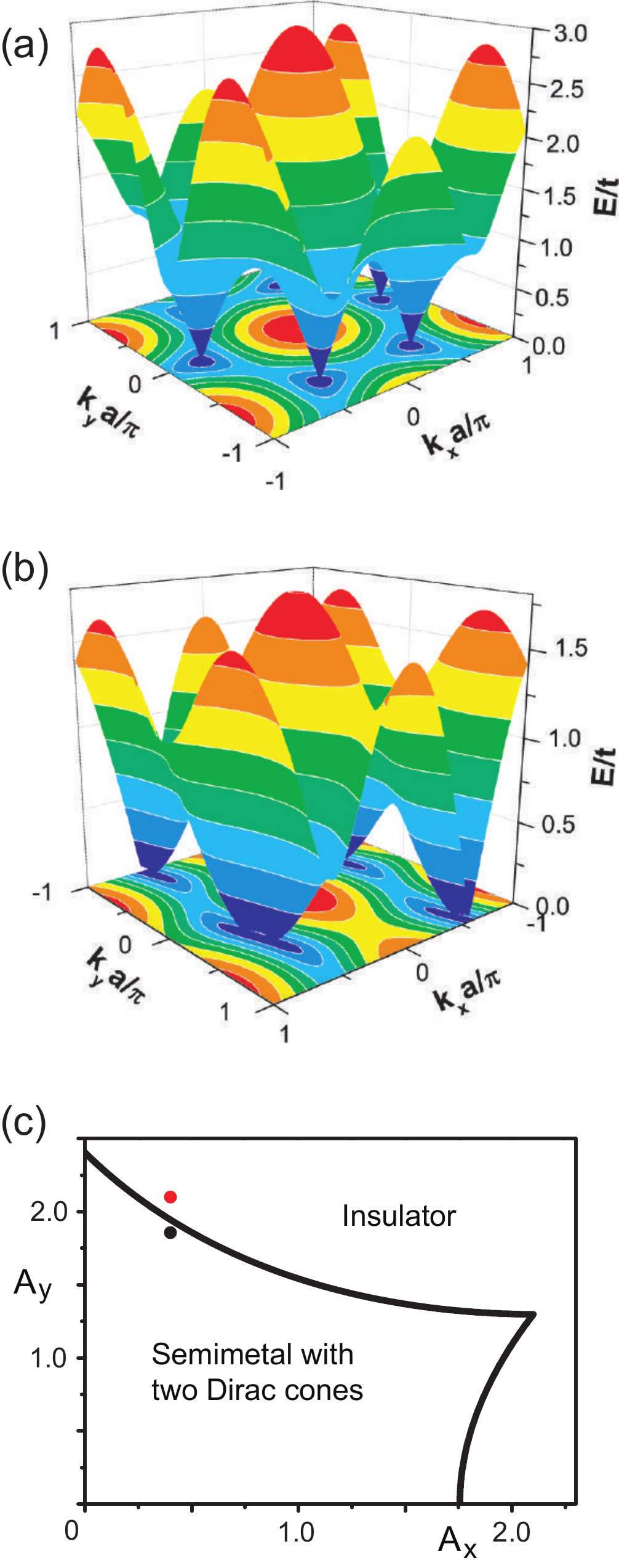}
\caption{(a) The tight-binding energy spectrum ($E>0$) of graphene and (b) with a driving $A_x=0.4$, $A_y=1.9$, $\varphi=\pi/6$. (c) The boundary between the semimetal phase and the insulator phase with merged Dirac points under the ac field, where the two dots are typical field strength under consideration.}\label{fig1}
\end{figure}

Since the Hamiltonian depends periodically in time, we employ results from the Floquet formalism in the following \cite{shirleyprb1965,selma12,gomez13,delplace13,goldman14,eckardt17}. Specifically, in the so-called high frequency regime $\hbar \omega_0 \gg  6t$, we neglect the coupling between different Floquet bands \cite{footnote1} and use a time-independent effective Hamiltonian description, which in the momentum space takes the familiar form
\beq
H_{eff}=-\sum_{\mathbf{k}}f^*(\mathbf{k})
a_{\mathbf{k}}^\dag b_{\mathbf{k}}+H.c.
\eeq
where $f(\mathbf{k})=t_{1} e^{i\mathbf{k}\cdot\delta_1}+t_{2} e^{i\mathbf{k}\cdot\delta_2}+t_{3} e^{i\mathbf{k}\cdot\delta_3}$ with \textit{dressed} anisotropic hopping $t_{m}=t\,J_0(\Gamma_{m})$. Here, $J_0$ is the 0-th order Bessel function of the first kind, $\Gamma_m=(e/\hbar)\,[(\mathcal{A}_{x} x_{m})^2+(\mathcal{A}_{y}
 y_{m})^2+2\mathcal{A}_{x}\mathcal{A}_{y}x_{m} y_{m}\cos\varphi]^{1/2}$
with $x_{m}\,(y_{m})$ denoting the $x(y)$ component of $\delta_m$. Throughout the paper, we restrict our study to the elliptical modulation $\varphi=\pi/6$, and choose $t$ as the energy unit and the dimensionless ac field strength $A_x=e\mathcal{A}_xa_0/\hbar$ and $A_y=e\mathcal{A}_ya_0/\hbar$.

The modified bandstructure is shown in Fig. \ref{fig1} as a function of the modulation amplitude $(\mathcal{A}_{x},\mathcal{A}_{y})$ (as the system is particle-hole symmetric, only the positive energy branch is shown). As seen in Fig. \ref{fig1}(b), while the bulk structure of the energy band away from the band center remains similar to the undriven case (Fig. \ref{fig1}(a)), the Dirac cones structure close to the charge neutrality point can reach, with a sufficiently strong ac field, the so-called Dirac cones merging scenario \cite{hasegawa06,montambauxprb2009,wunnjp2008}. The boundary between the semimetallic graphene and the merged scenario is shown in Fig. \ref{fig1}(c) \cite{delplace13}. While the physics of Dirac cones merging is originally formulated in the context of graphene, various analog systems \cite{naturetarruell2012,gomes12,prlbellec2013} have paved the way to its physical realization including new aspects being addressed \cite{polini13}. In the following, we will focus on the Hall transport phenomenon.

\section{Effect of ac field on the Hall conductivity}
We now apply a perpendicular magnetic field on the ac driven effective Hamiltonian to get
\begin{equation}
H=-\sum\limits_{i}\sum_{m=1}^3 t_m\, e^{i\theta_{i,\delta_m}}\,a_{i}^\dag\, b_{i+\delta_m}+H.c.
\label{hamil}
\end{equation}
The magnetic field is incorporated in the phase
$\theta_{i,\delta_m}=(e/\hbar)\int_i^{i+\delta_m}
\mathbf{A}_b(\mathbf{r})\cdot d\mathbf{r}$ with the vector potential
$\mathbf{A}_b(\mathbf{r})=\{-By,0\}$. The magnetic flux per hexagon is given by $\phi=3\sqrt{3}eBa_0^2/2\hbar$.
To obtain the Hall conductivity, we numerically evaluate the Kubo formula\cite{maprb2009}
\begin{equation}
\begin{array}{cll}
\sigma_{xy}
&=&\frac{i\hbar e^2}{S}
\sum\limits_{mn}\frac{f(\epsilon_m)-f(\epsilon_n)}{(\epsilon_n-\epsilon_m)(\epsilon_n-\epsilon_m+i\eta)}
 \langle n|v_x|m\rangle\langle m|v_y|n\rangle,
\end{array}\label{kubo}
\end{equation}
where $f(\epsilon)$ is the Fermi distribution function, $\epsilon_i$ is the eigenenergy corresponding to the state $|i\rangle$ for $i=m,n$, $S$ is the area of the sample, $v_x$ and $v_y$ are the velocity operators.
In the numerics, we consider a rectangular graphene sheet with width $N$ and the length $L$ (which contains
$N\times L$ carbon atoms) and take the temperature to be zero. In Sect. IV, we will in addition add onsite disorder terms $w_{i}$ to the Hamiltonian with amplitude randomly distributed in the interval $[-W/2,W/2]$ to study the effect of disorder. In this case, the Hall conductivity is averaged over up to $200$ random configurations.

\begin{figure}[tph]
\centering
\includegraphics[width=0.4\textwidth]{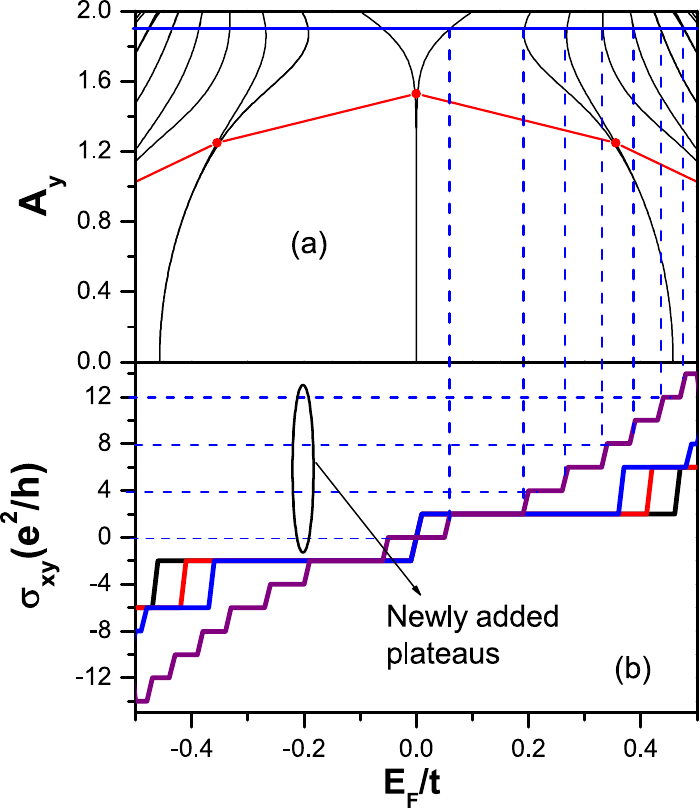}
\caption{(a) The Landau levels as a function of $A_y$ at fixed $A_x=0.4$. (b) The Hall conductivity curves as a function of the Fermi energy $E_F$ for different ac field strength: No driving (black); $A_y=0.8$ (red); $A_y=1.2$ (blue); $A_y=1.9$ (purple) at fixed $A_x=0.4$. Dash lines show the newly added integer plateaus.  The numerics are performed with $N=48$, $L=60$, and
$\phi=2\pi/48$. The behavior in the full energy range is shown in the Appendix A.\label{fig2}}
\end{figure}

\begin{figure}\begin{center}
\includegraphics[width=6cm]{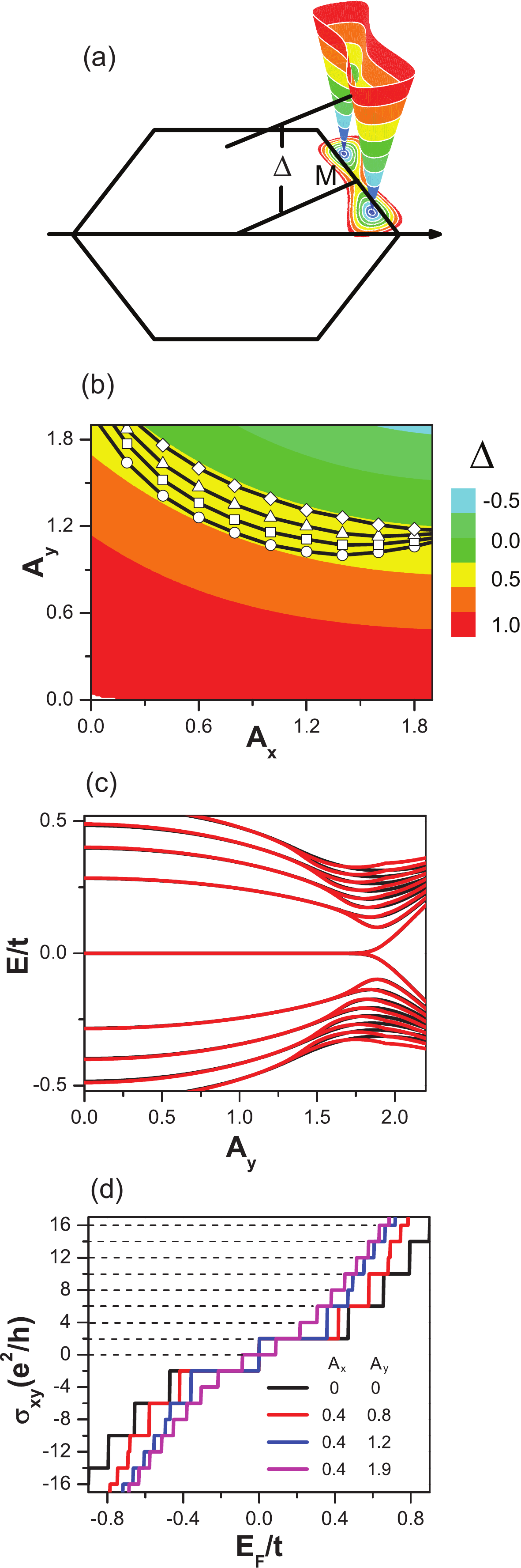}
\end{center}
\caption{(a) Energy spectrum of the low-energy Hamiltonian with two Dirac cones approaching each other at the high symmetry point $M$ in the Brillouin zone of graphene. The quantity $2|\Delta|$ denotes the energy gap at $M$.
(b) Contour plot of $\Delta$ as a function of
$(A_x,A_y)$. The various curves indicate the transition boundary
for $n=0$-$3$ Hall plateaus. (c) The LLs of the low-energy model as a function of $A_y$ at fixed $A_x=0.4$. The black curves are the tight-binding results. (d) The Hall conductivity obtained from the low-energy Hamiltonian for various ac field
strength. The other parameters are same as those in Fig. \ref{fig2}.}\label{fig5}
\end{figure}

\subsection{Quantum Hall plateaus}
The Hall conductivity is plotted in Fig. \ref{fig2}(b) as a function of the Fermi energy, around the band
center, for various $A_y$ values at fixed $A_x$ (red, blue and purple lines). For comparison, the black curve shows the Hall conductivity for the undriven graphene which exhibits the half-integer QHE with $\sigma_{xy}=\pm (n+1/2)(4e^2/h)$. 
In the presence of the ac field, we find that the effects are: (1) The width of the half-integer plateau shrinks; (2) As $A_y$ increases, new Hall plateaus appear at \textit{integer} values $\sigma_{xy}=\pm n (4e^2/h)$, in decreasing $|n|$ order as the ac field increases; (3) At sufficiently strong ac field, a Hall plateau eventually develops at $n=0$ (purple line in Fig. \ref{fig2}(b)); (4) Beyond this field strength, the full Hall conductivity sequence obeys $\sigma_{xy}=\pm n (2e^2/h)$ without the half-integer shift.

The appearance of new plateaus can be understood in terms of the evolution of the underlying Landau levels of the driven graphene. In Fig. \ref{fig2}(a) we show the corresponding Landau levels of the driven tight-binding model. In the undriven case $(A_y=0)$, the energy spectrum in the band center exhibits the well-known relativistic Landau levels due to the Dirac spectrum. With spin and valley degree-of-freedom, each LL is four-fold degenerate (in addition to the orbital degeneracy that is related to the sample size). As $A_y$ increases, each LL further splits into two sublevels, which indicates a degeneracy lifting. The order of splitting begins from higher to lower Landau levels (indicated in red dots).

Crucially, we find a one-to-one correspondence between the positions of the sublevel splitting and the development of new plateaus in the Hall conductivity. We also see that at fixed $A_y$ the separation of the two sublevels matches exactly with the width of the corresponding newly developed Hall plateau (indicated by blue dash lines). Since each splitting adds to the Hall conductivity by $2 e^2/h$, it shows that the effect of $A_y$ is indeed to lift the valley degeneracy, leaving with only factor two electron spin degeneracy. As for the extra half-integer shift in the Hall conductivity, we will address it in the next section.


\subsection{Low-energy effective model}
A further understanding on the development of new plateaus surrounding the band center can be gained by studying a low-energy model which describes the Dirac cones merging transition \cite{montambauxprb2009}, see Fig. \ref{fig5}(a).
The Hamiltonian is given by
\begin{equation}
\mathcal{H}=(\Delta-\frac{k_x^2}{2m} )\sigma_x+c_-k_y\sigma_y,
\label{radgrham}
\end{equation}
where $\sigma_i$ are the Pauli matrices acting on the pseudospin
space, $m$ and $c_-$ are $x$-direction band mass and $y$-direction Fermi velocity, respectively.
The merging parameter $\Delta$ drives the transition between a semimetallic phase with two Dirac cones $(\Delta>0)$ and an insulating phase $(\Delta<0)$, and at $\Delta=0$, the two Dirac points annihilate at the $M$ point in the momentum space.

In the presence of a magnetic field, the corresponding Landau level structure has been studied in Ref. \cite{dietlprl2008,montambauxprb2009}. For a fixed separation between the two Dirac cones $(\Delta>0)$, the Landau levels exhibit that of a parabolic band at high energies and a Dirac spectrum with valley degeneracy at low energies. For intermediate energies, the LLs are grouped into closely spaced sublevels, an effect due to the valley degeneracy lifting. Qualitatively speaking, the latter occurs when the magnetic length scale associated with the LL wavefunction becomes comparable to the inverse separation in the momentum space between the two Dirac cones \cite{montambauxprb2009}.

By comparing the energy spectra of the driven tight-binding Hamiltonian and the low-energy Hamiltonian, the parameters of the two Hamiltonians can be related, which are summarized in the Appendix B. Fig. \ref{fig5}(b) shows the contour plot of the merging parameter $\Delta$ as a function of the driving strength $(A_x,A_y)$. On the other hand,
the corresponding Landau levels are shown in Fig. \ref{fig5}(c) (red curves), which agree very well with the tight-binding calculations (black curves). The deviation at larger $A_y$ stems from the influence of the van Hove singularity in the tight-binding band spectrum, which moves towards the Dirac points with increasing $A_y$. Of course, the latter is not captured with the low-energy Hamiltonian.

\begin{figure}[tph]
\centering
\includegraphics[width=0.38\textwidth]{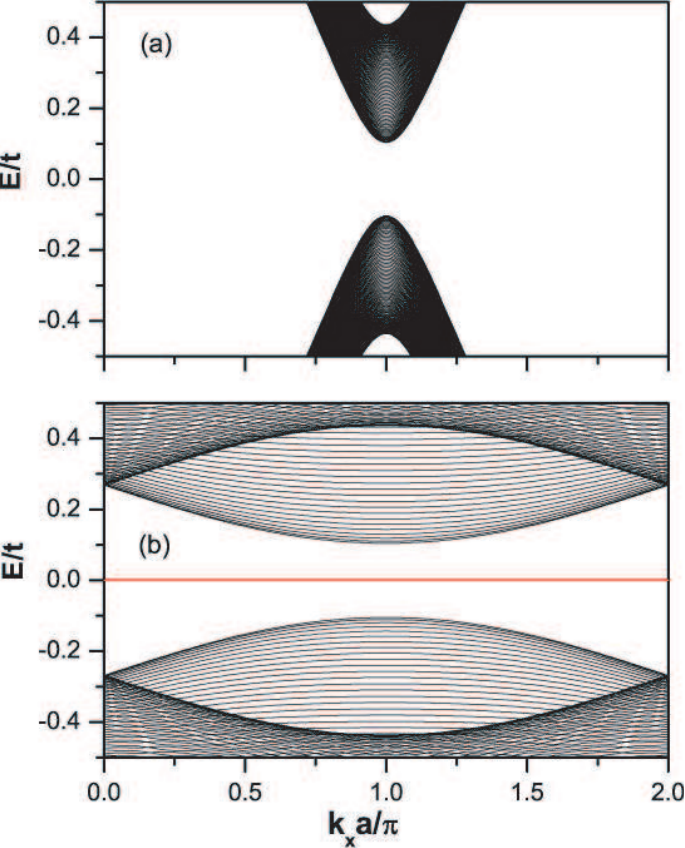}
\caption{(a) The bandstructure of graphene ribbons with zigzag ($N=96$) and (b) armchair edges ($N=232$) for $A_x=0.4, A_y=2.1$. The latter shows the emergence of dispersionless edge states in the gap.
\label{fig6-1}}
\end{figure}
In Fig. \ref{fig5}(d), the corresponding Hall conductivity is calculated with the Kubo formula using the numerically obtained eigenstates of the low-energy Hamiltonian. Comparing with Fig. \ref{fig2}(b), we also find a very good agreement between the two models. In Fig. \ref{fig5}(b), we show the transition boundary for the emergence of the new integer Hall plateaus (different symbols correspond to different plateaus). They show that the valley degeneracy lifting occurs already when the two Dirac cones come close together $\Delta \gtrsim 0$.

As for the disappearance of the half-integer shift in the Hall conductivity, the low-energy Hamiltonian provides a simple explanation in terms of the presence or absence of Berry phase for the electron cyclotron orbits in the momentum space \cite{montambauxprb2009}. In particular, when the Dirac cones are close together, depending on the energy of the cyclotron orbits, they can either enclose one or two Dirac cones of the opposite chirality. The two cases give rise to either a $\pi$ or zero Berry phase, respectively. For example, when two Dirac cones are sufficiently close, the lowest Landau level orbit always encloses two Dirac cones which results in no Berry phase, giving the $n=0$ plateau without the half-integer shift.

As a summary of this subsection, we show that the low-energy Hamiltonian for Dirac cones merging provides a quantitative physical picture, in the band center regime, of the emergence of new Hall plateaus in the driven graphene lattice.

\subsection{Edge states picture}
Here we numerically calculate the edge state bandstructure in a ribbon geometry (with zigzag and armchair termination on the edge) in the driven case with a magnetic field. This is to complement the previous results for a finite system.

Without the magnetic field, for the zigzag edge, we find a gapped edge spectrum at a sufficiently strong ac field, which is similar to the driven bulk graphene, see Fig. \ref{fig6-1}(a). For the armchair edge, however, a flat band traversing the whole Brillouin zone emerges in the strong ac field case (see Fig. \ref{fig6-1}(b)), which corresponds to states localized at the edge sample. This is reminiscent to the graphene ribbon with zigzag-bearded edges \cite{Wakabayashiprb2010}.

\begin{figure}[tph]
\centering
\includegraphics[width=0.46\textwidth]{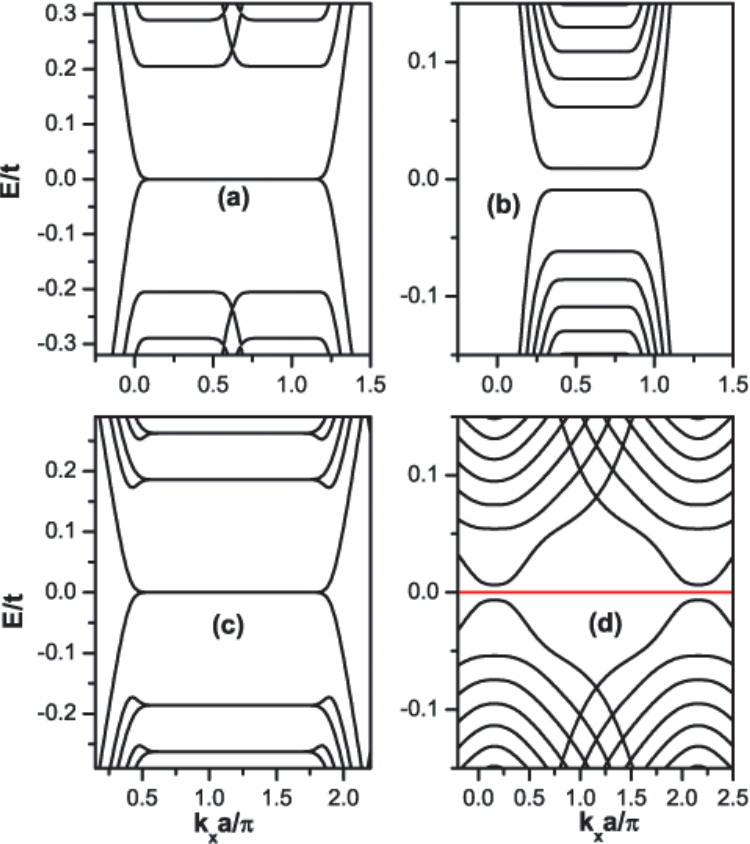}
\caption{The Landau levels of graphene ribbons with no driving (a,c) and (b,d) with driving $A_x=0.4, A_y=1.9$. (a,b) are for the zigzag edges ($N=96$, $\phi=2\pi/256$); (c,d) are for the
armchair edges ($N=262$, $\phi=2\pi/312$). 
\label{fig6-2}}
\end{figure}

With magnetic field but without the ac field, the subbands turn into
Landau levels \cite{breyprb,netoprb2006}, with Figs. \ref{fig6-2}(a) and \ref{fig6-2}(c) for the zigzag and armchair edges, respectively. With a strong ac field, we see that a gap eventually opens without dispersing edge states in both cases, see Figs. \ref{fig6-2}(b) and \ref{fig6-2}(d), showing a trivial insulating behavior at the charge neutrality point (no dispersive edge state crossing the Fermi level) - they provide an edge-state-picture understanding of the $n=0$ plateau.

As a side remark, for the armchair case, even though there are edge states at zero energy (red line in Fig. \ref{fig6-2}(d)), they are non-dispersive and thus they are not current carrying. Its origin is due to the boundary effect, as already seen in the case without a magnetic field in Fig. \ref{fig6-1}(b).

\begin{figure}[tph]
\centering
\includegraphics[width=0.7\textwidth]{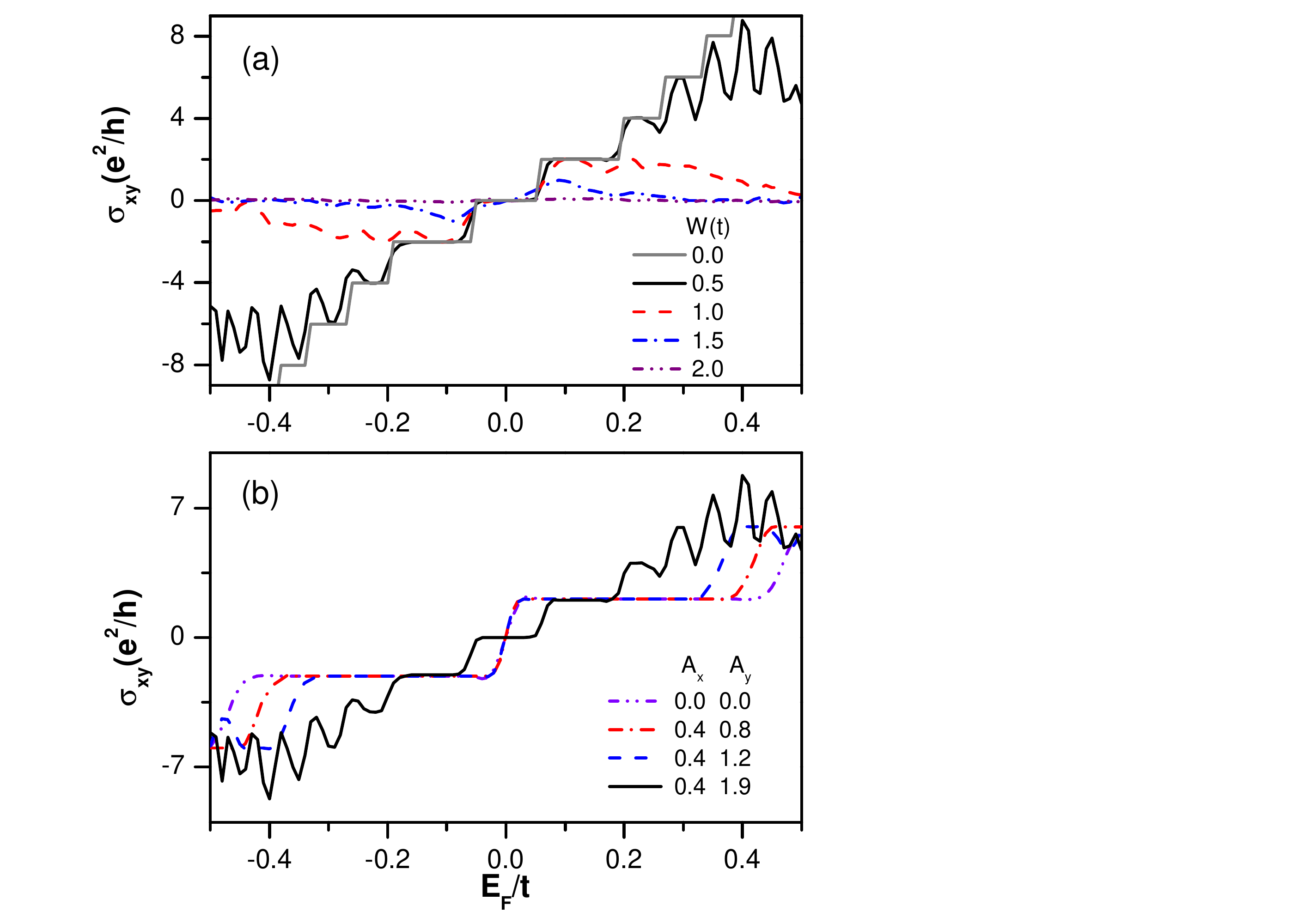}
\caption{(a) The Hall conductivity for different disorder strength at $A_x=0.4$ and
$A_y=1.9$, and (b) for the different ac field strength at $W=0.5t$. The other parameters are taken the same as those of Fig. \ref{fig2}.  \label{fig4}}
\end{figure}

\section{Effect of disorder}
We now turn to the  disorder effect on the QH plateaus in the driven graphene system. As illustrated in Fig. \ref{fig4}, the computed Hall conductivity indicates that the Hall plateaus shrink more rapidly away from the band center, which are destroyed in a one-by-one fashion with the increase of the disorder strength while the plateau near the band center is the last to vanish. Note that for a finite-sized sample (with $N=48$ and $L=60$ in Fig. \ref{fig4}), the numerical Hall conductivity still remains finite after the destruction of the QH plateaus and a finite-size scaling is needed in order to show that the electrons in the corresponding regimes become truly localized. The latter has been previously established \cite{shengprb2006} for the graphene system without the ac field, and the behavior shown in Fig. \ref{fig4} resembles the latter qualitatively. A similar trend of the destruction of the QH plateaus by disorder can also be computed based on the transfer matrix method \cite{shengprb2006,mkzp1983}, which is given in Appendix C [cf. Fig. \ref{fig7}].

The sequence of the destruction of the QH plateaus has been physically interpreted  \cite{shengprb2006} based on the flow of topological Chern numbers in the undriven graphene case, which should be applicable to the present case. 
Namely, the nonzero Chern numbers with opposite signs near the band edges [i.e., the region II in Fig. \ref{fig3}] move up/down towards the band center to annihilate those in the extended levels close to the zero energy, such that eventually a total annihilation of Chern numbers results in the destruction of the Hall plateaus \cite{shengprb2006,shengprl1997}. When the ac field is taken into account here, the LLs will further shift towards the zero energy while the interval between them is shortened in the absence of disorder (cf. Fig. \ref{fig3}). In turn, the extended levels separating different Hall plateaus merge together more easily in the presence of disorder, and eventually vanish due to the annihilation of the Chern numbers, thus speeding up the breakdown of the Hall plateaus, see Appendix C.

\section{Conclusions}

We studied the Hall transport properties of lattice model of graphene in the presence of an elliptically driven electric field. We used the Kubo formula to evaluate the Hall conductivity and showed that the half-integer QHE known in graphene can be transmuted into the integer QHE when the ac field is sufficiently strong. We showed that the ac field effectively modifies the low-energy part of the graphene tight-binding bandstructure, which can be captured within a low-energy Dirac cones merging Hamiltonian. The appearance of the new Hall plateaus can then be understood in terms of the gradual lifting of valley degeneracy in the Landau levels of Dirac cones systems. We also studied the evolution of the edge states, confirming the absence of topological edge current at the zero Hall plateau with the ac-field-driven integer QHE. In the presence of disorder, we showed that the ac field enhances the disorder-induced destruction of the quantum Hall states. The main qualitative features of the topological Chern numbers flow behaviour in the band center, however, is shown to be robust as in the undriven graphene.

\begin{acknowledgments}
Useful discussions with D. N. Sheng are acknowledged. The work was supported in part by the Scientific Research Fund of Hunan Provincial Education Department (Grant No. 13A109), the Natural Science Foundation of Hunan Province, China (Grant No. 2015JJ6005) (K.-H. D.), the Thousand Youth Talents Program of China (L.-K. L.), the NSFC (Grant No. 11474279), the MOST of China (Grant No. 2013CB933401), and the Strategic Priority Research Program of the Chinese Academy of Sciences (Grant No. XDB07010100) (G. S.), the NSFC (Grant No. 11534007), the MOST of China (Grant No. 2015CB921000), and National Key R$\&$D Program (2017YFA 0302902) (Z.-Y. W.).
\end{acknowledgments}

\appendix

\section{Hall conductivity in the full energy bandwidth}
\begin{figure}[tph]
\centering
\includegraphics[width=0.38\textwidth]{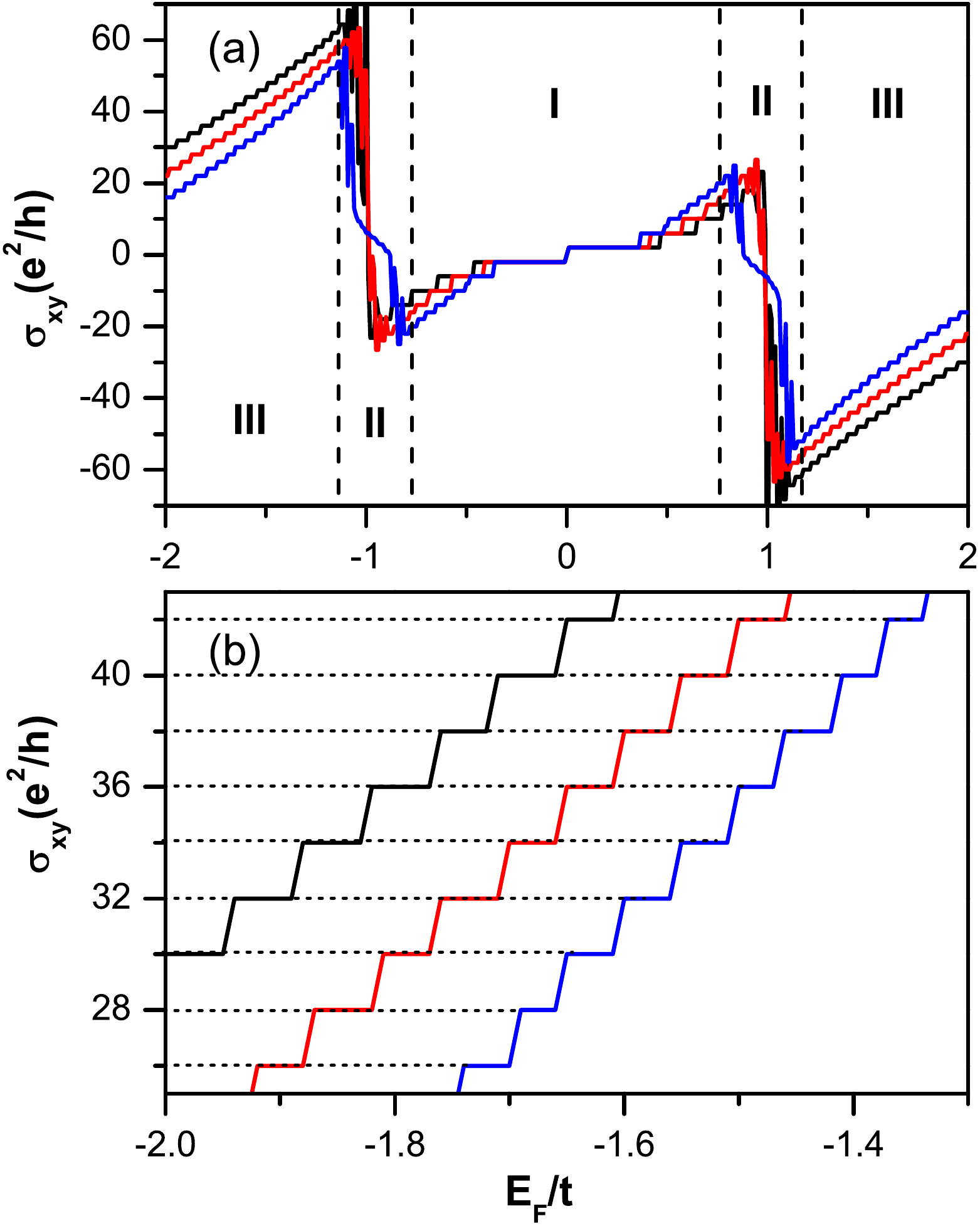}
\caption{(a) The Hall conductivity in the full energy band for different
ac field strength. (b) A close-up near the lower energy band edge. Black curve: $A_x=A_y=0$; Red curve: $A_x=0.4, A_y=0.8$; Blue curve: $A_x=0.4; A_y=1.2$. The other parameters are those of Fig. \ref{fig2}. 
\label{fig3}}
\end{figure}

The Hall conductivity in the full energy bandwidth is shown in Fig. \ref{fig3}(a) under the ac field. The result is qualitatively similar to the undriven case, as studied in Ref. \cite{shengprb2006}. Specifically, it features three characteristic regions: (I) band center regime; (II) van Hove regime; (III) band edge regime. In the presence of the ac field it is seen that while region (I) shrinks, region (II) is broadened. In region (III) the Hall plateau structure is essentially unaffected in the presence of $A_y$, see Fig. \ref{fig3}(b).

\section{Parameters fitting}
Here we give a precise fitting between the driven tight-binding Hamiltonian and the low-energy Hamiltonian parameters. For $\Delta
> 0$, $\Delta=-t_1+t_2+t_3$ the effective mass is $m=2\Delta/c_+^2$ with
\beq
c_\pm&=&\bigl[\frac{1}{2}(v_1^2+v_2^2+v_3^2)\nonumber\\
&&\pm\frac{1}{2}\sqrt{(2v_2v_3)^2+(v_1^2+v_2^2-v_3^2)^2}\bigr]^{1/2}
\eeq
where
$v_1=-\sqrt{3}\,t_1a\sin\lambda$, $v_2=(\sqrt{3}a/2)(t_3+2t_1\cos\lambda)$, $v_3=-3t_3\,a/2$ with
$\cos\lambda=(t_2^2-t_1^2-t_3^2)/(2t_1t_3)$.

For $\Delta \leq 0$, the effective mass satisfies
\beq
\frac{1}{m}=\alpha\cos^2\theta+\beta\sin^2\theta+\gamma\cos\theta\sin\theta
\eeq
where
$ \alpha =\frac{3}{4}a^2(-t_1+t_2)$, $ \beta =\frac{1}{4}a^2(-t_1+t_2+4t_3)$, $\gamma=-\frac{\sqrt{3}}{2}a^2(t_1+t_2)$, $ \sin\theta=\frac{v_x}{c_-}$, and $ \cos\theta=\frac{v_y}{c_-}$ with $ c_-=\sqrt{v_x^2+v_y^2}$, $ v_x= \frac{\sqrt{3}a}{2}(t_1+t_2)$, and $ v_y=-\frac{a}{2}(t_1-t_2+2t_3)$.

\section{Disorder effect }

\begin{figure}[tph]
\centering
\includegraphics[width=0.5\textwidth]{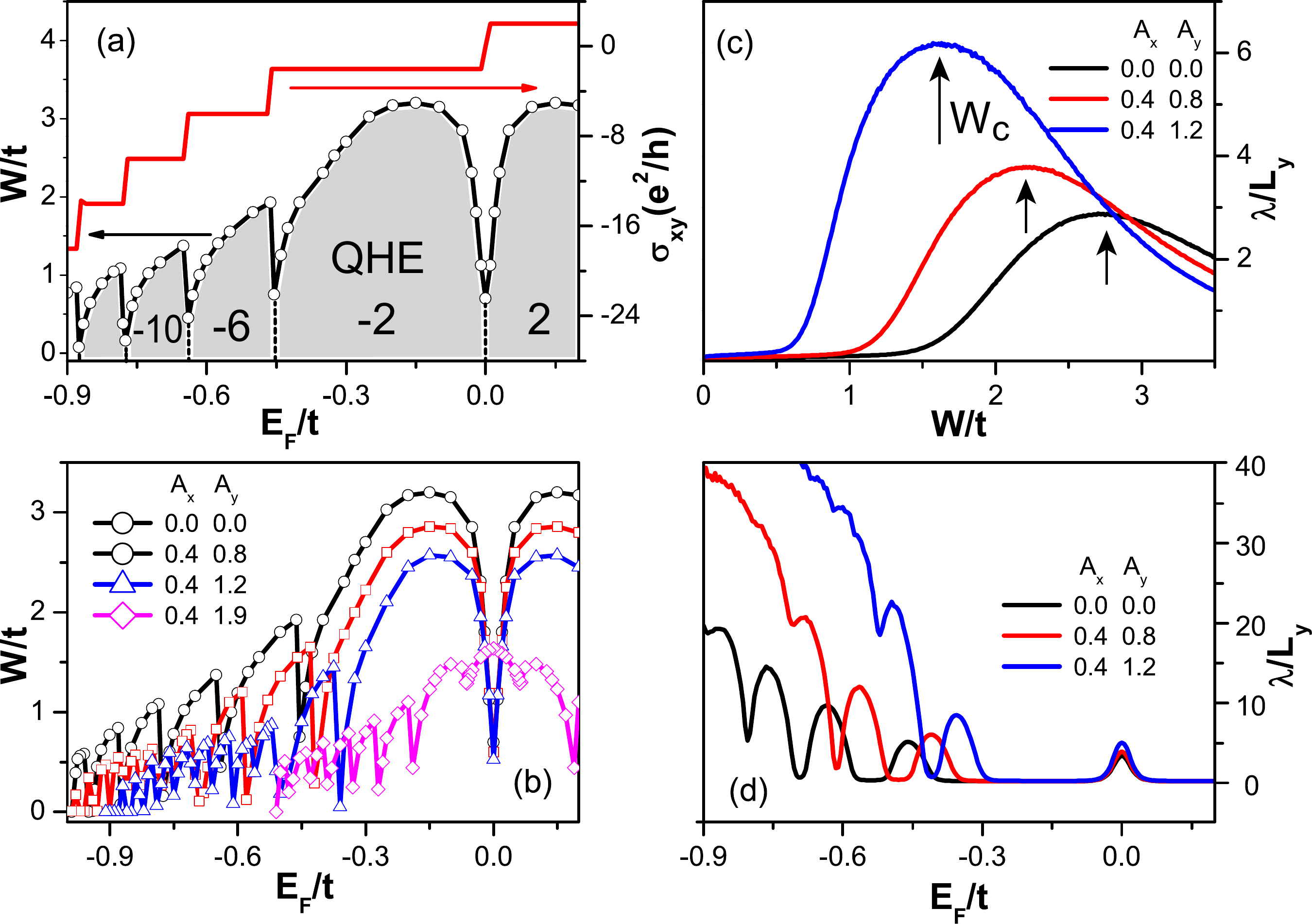}
\caption{(a) The phase boundary (circles) between the various quantum Hall states and the insulating regime in the undriven graphene for a finite-sized sample. The red curve indicates the corresponding Hall conductivity. (b) The variation of the phase diagram with the ac field strength. The normalized localization length, as a function of (c) the disorder strength $W$  at $E_F=-0.3t$; (d) the Fermi energy $E_F$ at $W=0.75t$, for various ac fields. The other parameters are taken the same as those of Fig. \ref{fig2}.
\label{fig7}}
\end{figure}

The disorder effect on the QH plateaus can also be analyzed based on the localization length computed using the transfer matrix method \cite{shengprb2006,mkzp1983}. Here we choose the sample as a bar with the width $L_y$ and length $L_x$, where $L_x$ is taken up to $10^{6}$.  We first summarize the effect of disorder on the phase diagram without the ac field in Fig. \ref{fig7}(a), which has been obtained previously in Ref.\cite{shengprb2006}. Note that the critical $W_c$ (circle) here is obtained by the peak of the localization length $\lambda$ (divided by $L_y$) in Fig. \ref{fig7}(c).  In principle, a finite size scaling is needed to decided $W_c$ in the thermodynamic limit \cite{shengprb2006,mkzp1983}, but here we fix $L_y=48$ and obtain the approximate results for the purpose of illustration. 

The effect of the ac field is to enhance the shrinking of the quantum Hall region with more dips developing inside the original plateau, see Fig. \ref{fig7}(b). With sufficiently strong ac field, the dip near zero energy evolves into a broad peak due to the appearance of the zero Hall plateau. Fig. \ref{fig7}(c) shows the localization length as a function of
$W$ for various ac field strength.  The sequential destruction of quantum Hall states is seen in Fig. \ref{fig7}(d).


\begin{thebibliography}{99}
\bibitem{footnote1} The derivation of the effective Hamiltonian within the Floquet formalism can be found, for e.g., in Refs. \cite{delplace13,goldman14,eckardt17}, which we do not repeat here.
\end{thebibliography}
 \end{document}